\documentclass[aps,prd,11pt,floatfix,groupedaddress]{revtex4}
\usepackage[dvips,final]{graphicx}
\usepackage{amsmath}
\usepackage{amssymb}
\usepackage{url}
\usepackage{subfigure}
\usepackage{textcomp}

\begin{document}

\title{Level Crossing Analysis of Cosmic Microwave Background Radiation:\\ A method for detecting cosmic strings}

\author{M. Sadegh Movahed$^{1}$}
\email[]{m.s.movahed@ipm.ir}
\author{Shahram Khosravi$^{2,3}$}
\email[]{khosravi@ipm.ir}

\affiliation{$^{1}$Department of Physics, Shahid Beheshti University, G.C., Evin, Tehran 19839, Iran}
\affiliation{$^{2}$Department of Physics, Faculty of Science, Tarbiat Moallem University, Tehran, Iran}
\affiliation{$^{3}$School of Astronomy, Institute for Research in Fundamental Sciences, (IPM), P. O. Box 19395-5531, Tehran, Iran}

\date{\today}

\begin{abstract}
In this paper we study the footprint of cosmic string as the
topological defects in the very early universe on the cosmic
microwave background radiation. We develop the method of level
crossing analysis in the context of  the well-known Kaiser-Stebbins
phenomenon for exploring the signature of cosmic strings.  We
simulate a Gaussian map by using the best fit parameter given by
WMAP-7 and then superimpose cosmic strings effects on it as an
incoherent and active fluctuations. In order to investigate the capability of
our method to detect the cosmic strings for the various values of
tension, $G\mu$, a simulated pure Gaussian map is compared with that of including cosmic strings. Based on the level crossing analysis,
the superimposed  cosmic string with $G\mu\gtrsim 4\times 10^{-9}$
in the simulated  map without instrumental noise and the resolution
$R=1'$ could be detected. In the presence of anticipated
instrumental noise the lower bound increases just up to $G\mu\gtrsim
5.8\times 10^{-9}$.
\end{abstract}

\maketitle

\section{Introduction}

One of the most interesting predictions of quantum field theory in
the domain of cosmology is the possibility of transition between
different vacuum states during the expansion of the universe. Depending
upon the topology of these states, a series of stable topological
defects such as domain walls, monopoles and cosmic strings can be
formed. Regarding the formation of strings, they may have
self-interactions causing the formation of closed loops according to the 
so-called inter-commutation
\cite{neil07,neil10,matt09,hind94,sak06,vak84,vil85,vilinkin00,shell87,maxim99}.  In
principle, string loops start to oscillate and emit stochastic
gravitational waves which results in their annihilation,
\cite{allen96} while the infinite straight strings survive up to
now. Although, in some models of inflation such as false vacuum
dominated inflation (in the context of hybrid inflation theories),
topological defects are formed at the end of inflation era
\cite{cope94}, but the true focus on the issue of cosmic strings was in the recent years, mainly
because new models of inflation derived from the superstring theory,
result in acceptable possibility for the production of cosmic
strings \cite{sar02,tye06,kibb04,cop03}. This is based on the consideration of cosmic string networks
consisting of infinite strings, loops, and junctions of two or more
strings which will definitely have effects such as lensing, CMB
polarization, and of course CMB anisotropies. Astrophysical evidence
of cosmic string strongly depends on the two following parameters,
namely: dimensionless string tension $G\mu/c^2\sim
\Lambda^2/M^2_{\rm Planck}$ ($G$ is the Newton's constant and
$\Lambda$ stands for energy scale when the strings are created) and
inter-commuting probability, $P$. The quantity $\mu$ shows the mass
per unit length of cosmic string. The energy density of cosmic
string is associated with the scale of phase transition and symmetry
breaking in which cosmic strings are produced. Generally the formation of cosmic strings could occur at very
extended ranges of energy scale, e.g. the Grand Unified Theory (GUT)
scale with $\Lambda\sim 10^{16}$GeV  corresponding to $G\mu/c^2\sim
10^{-6}$, consequently there is a wide range of $G\mu/c^2$s has
been supposed \cite{vilinkin00,firouz05}.
It must
be noted that the footprints of strings, such as anisotropies in CMB
are directly affected by their tension. So determining the bounds
for the tension, directly means limiting the fundamental theory on
the basis of which cosmic strings are produced. To this end, observing the cosmic string footprints by using various approaches
can be interpreted as a kind of observational evidence for the low
energy limit of the superstring theory and would provide the most
direct test of string theory and could rule out or constrain on
particle physics models.

From theoretical and observational perspectives, there are a dozen
constrains on the cosmic string's parameters. Recent analysis by
using pulsar timing and photometry based on gravitational
microlensing put a constraint for tension in the range
$10^{-15}<G\mu/c^2<10^{-8}$ \cite{jent06,psh09,psh10,dam05,rich10,okn99,khlop89, khlop86,khlop82}.
Using COSMOS survey, there is  $G\mu/c^2<3\times 10^{-7}$  for cosmic string reported in \cite{smoot10}. 
The $21$ $cm$ signature of cosmic string wakes has also been explained in Ref. \cite{branddd10}.
Another robust constraint on the cosmic string's free parameters
comes from the temperature fluctuations at the last scattering
surface. More recent full analysis and prediction for incoming
satellite based surveys can be found in
\cite{neil07,neil10,reg09,mark09}. Temperature fluctuations contain
the accumulation of anisotropies induced by cosmic strings and can
be divided into two categories: 1) anisotropies created by the
so-called Kaiser-Stebbins effect and those related to
pre-recombination processes and 2) the stochastic background of
gravitational waves produced by decaying of string loops
\cite{kaiser84,spergel08}. Calculation of temperature angular power
spectrum  puts the upper bounds  $G\mu/c^2<2\times 10^{-7}$\cite{spergel08,pre93b,bevis05,wyman05,wyman06,bevis07,fra05,kaiser84}
and $G\mu/c^2<6.4\times 10^{-7}$ at $2\sigma$ confidence interval
for Abelian-Higgs case \cite{rich10}. Nambu-Goto numerical simulation for network of cosmic string has been used by A. A. Fraisse et al. \cite{spergel08}.  For detecting the anisotropies induced by mentioned strings, temperature gradient magnitude operator has been applied in this paper.  Ligo and Virgo collaborations
recently have reported the newly lower and upper bounds  as $7\times
10^{-9}< G\mu/c^2 < 1.5\times 10^{-7}$ on the stochastic
gravitational waves produced by cosmic strings \cite{ligo09}. The
skewness in CMB fluctuations and its dependency on inter-commuting
probability has been investigated in \cite{daisu10}. Based on B-mode polarization of cosmic microwave background new constraint on $G\mu$ has been reported in Ref. \cite{zhe10}. 

Direct implication based on the explicit recognition of
discontinuity in the fluctuation of cosmic microwave background
radiation  is a unique signature of straight cosmic string , namely
the Kaiser-Stebbins effect \cite{kaiser84}. In this part, there are also some upper bounds as well as
lower bounds for distinguishing cosmic strings. Another method is the so-called Wavelet domain Bayesian denoising which has recently been used by  D. K. Hammond et al. \cite{hammond09}. Since the cosmic string is characterized by a localized edge like discontinuity, they used a Steerable Pyramid wavelet  transform to discriminate straight edge from other features. Transformation has been done with 6 orientations and 4 spatial scales.  According to their results, 
the signature of cosmic strings in the  CMB map without noise can be
identified for tensions in the range $G\mu/c^2\gtrsim6.3\times
10^{-10}$ while for a noisy map the lower bound increases more than
one order of magnitude \cite{hammond09}. For completing this topic, reader can also refer to Refs. \cite{barr01,barr06}.  The detectability threshold for cosmic string signal  in the presence of anticipated noise such as extra-Galactic and Galactic foregrounds has been studied  in  \cite{jeong10}.  A series of papers published
by Brandenberger and his collaborators have been devoted to the
ability of so-called Canny algorithm to detect the cosmic string and
super strings according to Kaiser-Stebbins effect
\cite{brand081,brand080,brand09}. They have concluded the lower
bound of $G\mu/c^2\gtrsim 5.5\times 10^{-8}$.  Actually in the Canny method, the sharp and strength edges are picked up  by means of the magnitude of the contrast  from one side of produced edge to the other one \cite{brand081}.  Non-gaussianity due
to cosmic strings in addition to other events such as kinetic
Sunyaev-Zel'dovich effect have been examined in \cite{stark03} by
multi-scale methods (also see \cite{chi10} for more discussions). They have argued that to find a significant
results not only one should use robust methods but also a
combination of powerful methods is necessary.
\begin{figure}
\begin{center}
\includegraphics[width=0.4\linewidth]{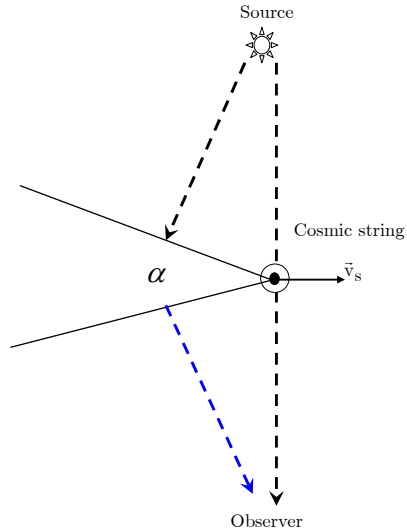}
\caption{\label{ks} A cross-section of space-time perpendicular to a straight cosmic string. Due to deficit angle, the light ray passes through the left part of plot to be blueshifted with respect to that of received from the right part.}
\end{center}
\end{figure}

Another most interesting method which is potentially able to discriminate map with and without cosmic strings as well as other topological defects is the so-called peak-peak correlation function \cite{peac85,bardeen86,hevan99,hevan01}.
In this method,  two-point correlation function of the local maxima (or minima) of temperature fluctuations on the CMB map at a typical peak threshold is determined. A. F. Heavens et al. have demonstrated that this correlation function for a simulated CMB map accompanying cosmic string grows up significantly at around $10-15$ arcminutes, relative to the pure Gaussian map. As they claimed, this procedure has no adjustable parameter, consequently it can be powerful for test of Gaussianity hypothesis as well as tracking the footprint of other mechanism for producing initial fluctuations in the upcoming high resolution observations. Although, they didn't show the efficiency of this method for various values of $G\mu$ which is the main purpose of  this paper, but it is interesting reapply the same idea based on hotspots (or coldspots) framework.  In our method not only we investigate all features of fluctuations instead of local extrema used in the peak-peak correlation function, but also the $q-$moments of relevant quantity are used (see section II for more details).  

In this study we are relying on a newly demonstrated method to put a
lower limit on the tension of straight cosmic strings in the high
resolution of CMB map for which we can discriminate their
signatures. 
As the effect of cosmic strings, we concentrate on the
discontinuities and fluctuations in the CMB map arising from the Kaiser-Stebbins
effect which is a part of integrated Sachs-Wolfe effect and
 can produce observational consequences on the anisotropies in the CMB
 map. This feature is based on the gravitational lensing which was a well-established phenomena
 by the time \cite{kaiser84,stebb88,birkin83,aghanim98,Birk89}. 
 Consider a string extended in the direction
 perpendicular to the paper (see Figure (\ref{ks})), and two photons emitted from
 a source in the background. Choose two paths on opposite sides of
 the string that reach the same observer. If the string is moving in
 the direction perpendicular both to its length and to the line of
 sight between the source and observer, with the velocity $\textbf{v}_s$, a
frequency shift will occur between two photons. Thus if we move
 across the cosmic string when sweeping the
CMB we can see a jump in the anisotropy map. The amount of this jump
is
 \begin{equation}\label{kseq1}
 \frac{\delta T}{T}=8\pi G\mu\, |\hat{\textbf{n}}.(\gamma_s \textbf{v}_s\times
 \hat{\textbf{e}}_s)|
 \end{equation}
 in which $\hat{\textbf{n}}$ is the direction of observation,
 $\textbf{v}_s$ is the velocity of the string, $\hat{\textbf{e}}_s$
 its orientation, and $\gamma_s$ is the Lorentz factor for the
 string. Eq.(\ref{kseq1}) can provide a basis for detecting cosmic strings via direct
 observation of the CMB anisotropy map. However, it must be noted
 that since the Kaiser-Stebbins effect may be considerably
  smaller than the observed temperature fluctuations in WMAP data,
  we cannot expect
 to detect strings in the WMAP data. Meanwhile we are able to
 produce simulated maps of CMB with any set of parameters. So in
 order to examine the bounds on the string tension imposed from this
 analysis, we can use simulated maps to apply certain
 methods of statistical analysis on the inhomogeneities to detect the
 strings or to assume their existence and then obtain better bounds
 on string tension. 

The rest of paper is organized as follows:  In Sec. II we introduce
level crossing analysis to investigate the fluctuation of
temperature on the last scattering surface with and without cosmic
strings. Simulation maps of CMB for various values of so-called
string tension using most recent observation based on WMAP-7 mission
will be given in details in Sec. III. Data analysis by using a robust
method in complex system from a statistical point of view to
distinguish the Gaussian map with an without cosmic strings in the
presence of expected instrumental noise will be explored in Sec. IV.
Summary and  conclusions are presented in Sec. V.
\begin{figure}
\begin{center}
\includegraphics[width=0.5\linewidth]{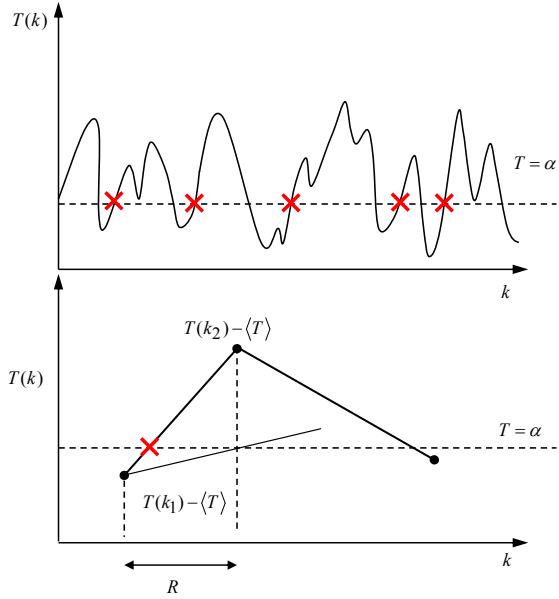}
\caption{\label{level1} Positive slope crossing at the level $T=\alpha$.}
\end{center}
\end{figure}
\section{Level crossing analysis on the flat sky map}
After the innovation of Rice for calculating the upcrossings and
downcrossings of a stochastic signal at an arbitrary level
\cite{rice44}, many studies have been devoted to signal processing
as well as fluctuations of height of rough surfaces from the level
crossing point of view as a powerful method in complex system
\cite{percy00,movahed06,movahed05,movahed07,tabar02,tabar03}.
In the level crossing analysis, we are interested in determining the
number crossings of the temperature fluctuations, $N_{\alpha}$, at
an arbitrary level $\alpha$. We consider a sample
function of temperature fluctuations which is defined on the
homogeneous and isotropic random surface represented  by
$T(\hat{n})$ at direction $\hat{n}$ relative to the observer placed
at the center of sphere.
For a flat patch of CMB map, we assign temperature fluctuations to each point by $T(x_i,y_i)$, where $x_i$ and $y_i$ demonstrate the coordinate positions. 
For convenience, the origin of the coordinate system of the patch is
placed in the left bottom of corresponding map.

Since we consider the statistical isotropy and homogeneity of
temperature random field on the surface,  without losing generality,
we assume  a one-dimensional slice of temperature fluctuations on a
patch of CMB with size $\Theta$ parallel to $x$ or $y$ axes. The
size of this signal depends on the resolution of data set which is
simulated or observed depicted by $R=N/\Theta$. Also the numbers of
pixels will be  $N^2$. Now consider a sample of an ensemble of one
dimensional signal of temperature fluctuation, $T(k)$, for which $k$
runs from $1$ to $N$. Suppose  $n_{\alpha}^{+}$ denotes the numbers
of positive slope crossings (upcrossings) of, $T(k)- \langle T
\rangle = \alpha$, for a typical sample size $\Theta=R\times N$ (see
Figure (\ref{level1}) ). The ensemble averaging for level crossing
with positive slope is also given by:
\begin{equation}\label{ensemble}
N_{\alpha}^{+}(\Theta)=\langle n_{\alpha}^{+}(\Theta)\rangle.
\end{equation}
For a statistical isotropic and homogenous fluctuation, the ensemble averaging can be done on various one dimensional slices of fluctuations. For convenience, we chose these slices to be parallel to  $x$ or $y$ axes. Furthermore,  the average number of crossings is proportional to the space (time)
interval $\Theta$ \cite{percy00}. Hence:
\begin{eqnarray}
N_{\alpha}^{+}(\Theta)&\propto& \Theta\nonumber \\
N_{\alpha}^{+}(\Theta)&=&\nu^{+}_{\alpha} \Theta
\end{eqnarray}
which $\nu_{\alpha}^{+}$ is the average frequency of positive slope
crossing of the level $T(k) - \langle T \rangle=\alpha$.  It must be pointed out that in the long run processes the conservation law for upcrossings and downcrossings will be satisfied \cite{percy00}. In our analysis due to the statistical isotropy and 200 numbers of run over ensembles or even more, this conservation law statistically holds and so, we use the upcrossing events throughout our analysis.
\begin{figure}
\begin{center}
\includegraphics[width=0.5\linewidth]{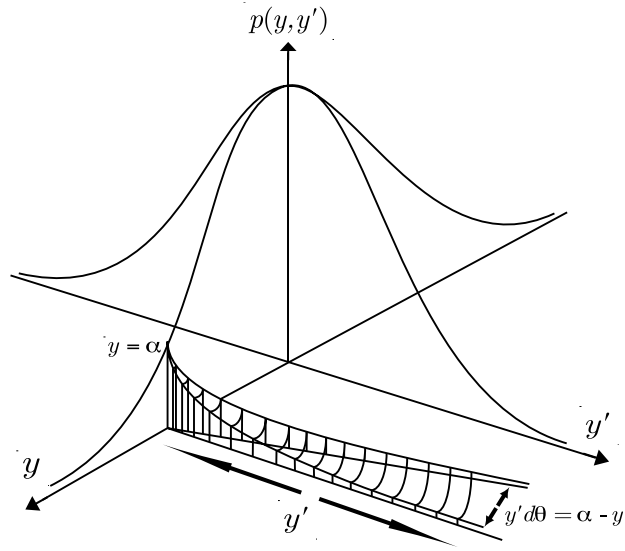}
\caption{\label{proba} Sketch of joint probability density function of a typical
fluctuation and its derivative with respect to corresponding dynamical parameter (angle)
in the level crossing theory. The shaded area indicates the total probability of crossing with positive slope at level $y=\alpha$.}
\end{center}
\end{figure}
From mathematical and statistical points of view, the frequency parameter $\nu_{\alpha}^{+}$ can be
deduced from the probability distributions associated with the temperature  fluctuations,
 $T(k) - \langle T \rangle $.  In order to calculate this probability distribution function,  we assume two necessary and sufficient conditions for crossing the level $T=\alpha$ by the underlying signal with positive slope as follows:\\
1) At the beginning of the interval we should have $T(k_1) - \langle T
\rangle < \alpha$.\\
2)  The slope of signals should be larger or at least equal to the
slope of line which is drawn between the point at the beginning of
interval and the point located on the horizontal line of level
$\alpha$ (Figure (\ref{level1})), namely:
\begin{equation}
 \frac{d \left [T(k) -
\langle T \rangle \right]}{d\Theta}> \frac{\alpha-\left [T(k_1)
- \langle T \rangle \right] }{d\Theta}
\end{equation}
where $d\Theta=R$. If the above-mentioned conditions are satisfied
\cite{percy00,movahed06,movahed05,movahed07,tabar02,tabar03},
we statistically expect to get a high probability for a crossing in the interval $d\Theta$. In order to determine whether
the above conditions are satisfied at any arbitrary location $k$, we
should find how the values of $y\equiv T(k)-\langle T \rangle $ and
$ y ^{\prime}\equiv \frac{ dy }{d\Theta}$ are distributed by
considering their joint probability density function,
$p(y,{y}^{\prime})$. For a specific level $y=\alpha$ and interval
$d\Theta$, we focus on the values of $y < \alpha$ and values of
${y}^{\prime}=(\frac{dy}{d\Theta}) > \frac{\alpha-y}{d\Theta}$,
corresponding to  the region between the lines $y=\alpha$ and
${y}^{\prime}= \frac{\alpha-y}{d\Theta}$ in the plane
($y,{y}^{\prime}$) (see Figure (\ref{proba})). Subsequently, the
probability of positive slope crossing of signal at level $y=\alpha$
in the $d\Theta$ interval is given by:
\begin{eqnarray}\label{1}
{\rm Probability}=\int _0^{\infty} dy ^{\prime}\int _{\alpha-y^{\prime}d\Theta}^{\alpha}p(y,y^{\prime}) dy\end{eqnarray}
 When $d\Theta\rightarrow 0$, it is legitimate to put:
\begin{equation}
p(y,{y}^{\prime})=p(y=\alpha,{y}^{\prime})
\end{equation}
Since at large values of $y$ and ${y}^{\prime}$, the probability
density function approaches zero fast enough, Eq.(\ref{1})
may be written as:
\begin{equation}
{\rm Probability}= \int_{0}^{\infty}
d{y}^{\prime}\int_{\alpha-{y}^{\prime}d\Theta}^{\alpha}
p(y=\alpha,{y}^{\prime})dy
\end{equation}
In this form,  the integrand does not depend on $y$ so the first
integral can be integrated easily:
\begin{equation}
{\rm Probability}=d\Theta \int_0^{\infty} p(y=\alpha,{y}^{\prime}){y}^{\prime}dy^{\prime}
\end{equation}
Figure (\ref{proba}) shows the area of interest for probability of
upcrossings at level $\alpha$.  As mentioned before, the average
number of crossing with positive slope in interval $\Theta$ is
$\nu^{+}_{\alpha}\Theta$. So the average number of positive
crossings of $y=\alpha$ in interval $d\Theta$ is equal to the
probability of positive crossings of $y=\alpha$ in $d\Theta$.  We can write the
average number of upcrossings at the level $\alpha$ in terms of the
joint probability density function as follows:
\begin{equation}
\nu_{\alpha}^{+}d\Theta=d\Theta \int_{0}^{\infty}p(\alpha,{y}^{\prime}){y}^{\prime}d{y}
^{\prime}
\end{equation}
or
\begin{equation}
\nu_{\alpha}^{+}=\int_{0}^{\infty}p(\alpha,{y}^{\prime}){y}^{\prime}d{y}
^{\prime}
\end{equation}
Another useful parameter based on $\nu_{\alpha}^{+}$ can be introduced as:
 \begin{equation}\label{ntq}
N_{tot}^{+}(q)=\int_{-\infty}^{+\infty}\nu_{\alpha}^{+} |\alpha -
\bar{\alpha}|^{q} d \alpha.
\end{equation}
\begin{figure}
\begin{center}
\subfigure{\includegraphics[width=0.5\linewidth]{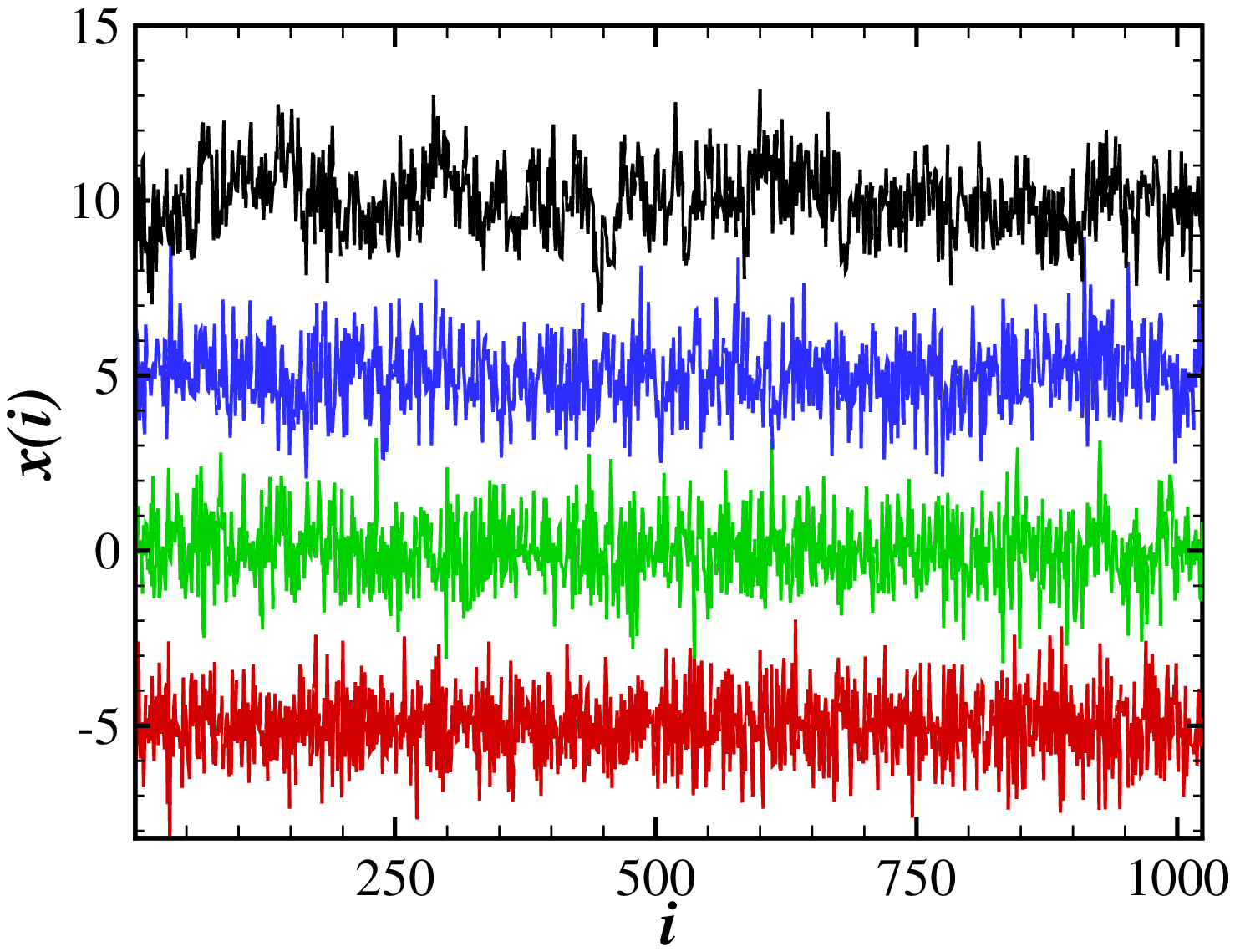}}
\subfigure{\includegraphics[width=0.5\linewidth]{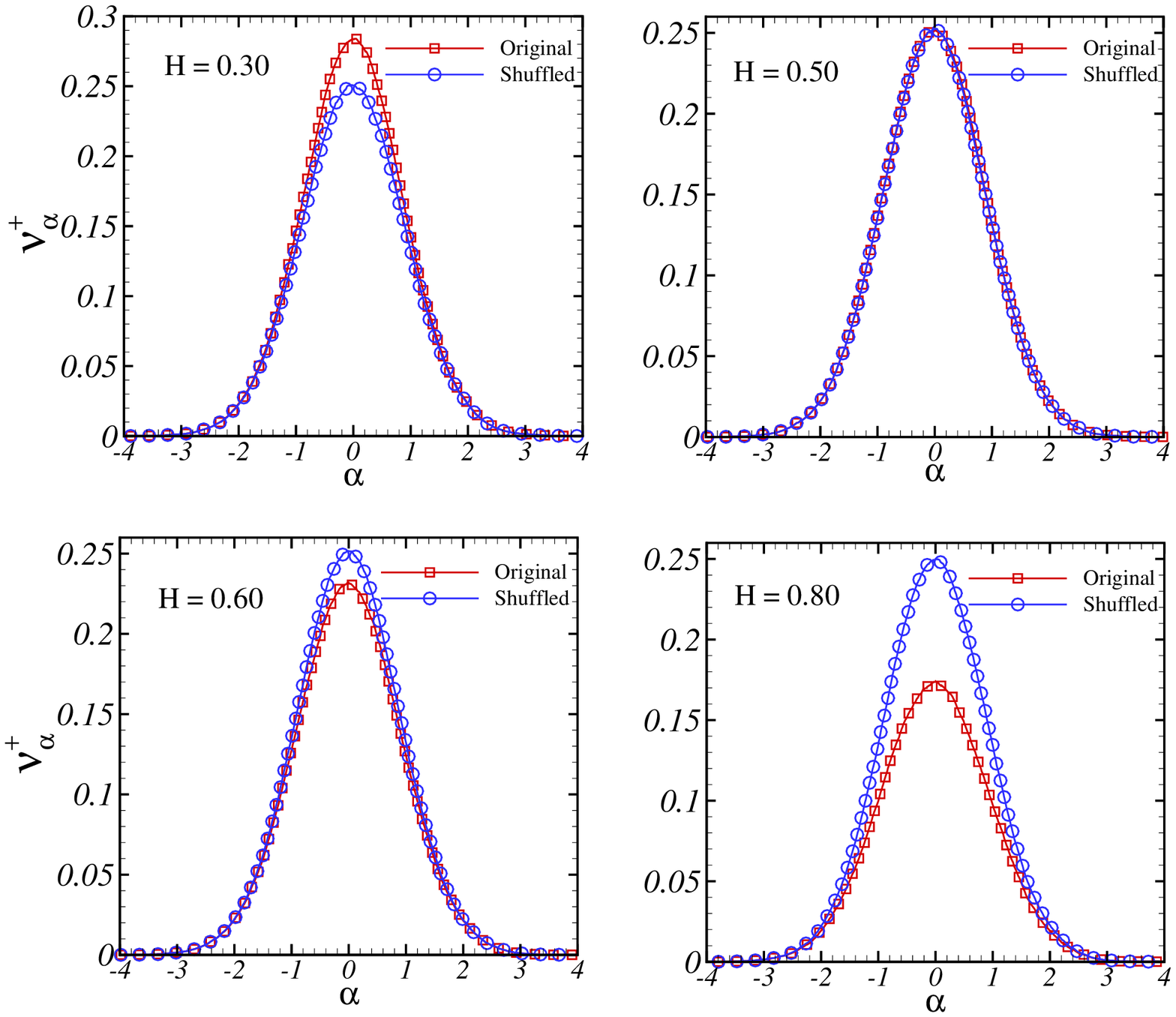}}
\caption{\label{hurst1} Upper panel corresponds to generated series for various values of roughness exponent (from top to bottom: $H=0.8$, $H=0.6$, $H=0.5$ and $H=0.3$). Lower panel shows the level crossing analysis of time series with above mentioned Hurst exponent.  Results regarding the original and shuffled series have been indicated in each panel.} \end{center}
\end{figure}
Obviously, for $q=0$ the above quantity determines total number of
upcrossings for temperature fluctuations with positive slope at all
levels. For a typical rough fluctuation, $N_{tot}^{+}(q=0)$ is
larger than that of the smooth signal and consequently, this quantity
is a criterion for the roughness of the processes. For a correlated
signal, $N_{tot}^{+}(q=0)$ is smaller than that of a completely
random one, while for  anti-correlated data set $N_{tot}^{+}(q=0)$
is larger than the same series for which its memories have been
destroyed by shuffling. Figure (\ref{hurst1}) shows the level
crossing results for some fractional Gaussian noises (fGn) and  the
corresponding shuffled signals. For an anti-correlated  series  the
so-called Hurst exponent is $H<0.5$, for a completely random
Gaussian data set  $H=0.5$ and for a correlated signal $H>0.5$
\cite{bunde,movahedsun}. In addition,  for moments $q<1$, those
terms in $N_{tot}^{+}(q<1)$ will become dominant that have small
deviations from mean level, $\bar{\alpha}$, which demonstrate the
statistics of small fluctuations. On the contrary, for $q>1$, those
terms with large fluctuations in the integrand of Eq.(\ref{ntq}) are dominant in $N_{tot}^{+}(q>1)$  explaining the statistical
properties of the upcrossings far from mean level. These terms
correspond to the rare events. 

In what follows, we are going to derive
$\nu_{\alpha}^{+}$ for discrete temperature fluctuations in the simulated map with and without cosmic strings on the flat sky and compare them to find a robust criterion to distinguish pure Gaussian fluctuations from fluctuations containing cosmic string components. The motivation of using this simple method can be justified according to the following reasons:\\
i) Since the statistical isotropy is valid as a major statistical property \cite{amir1,amir2,sadeghaniso}, one can cut many
one-dimensional signals in every directions and  by ensemble averaging, compute the frequency of upcrossings (downcrossings) at all
interested levels along with their variances.  \\
ii) As one can see in the next section, the superposition of fluctuations produced by
cosmic strings can generate new and extra ups and downs in temperature fluctuations, consequently finding such
 statistically meaningful footprints in the map in comparison with pure Gaussian signature including instrumental noise may potentially help
 us to get deep insight in the cosmic string detections.  \\
iii) Due to the  phase coefficient  in the Fourier analysis, it seems that many trivial imprints of cosmic strings
 diminish or at least are mixed with other observational  phenomena so it is another motivation to investigate the imprint of cosmic strings in the real space as mentioned in Ref. \cite{brand080}.\\
\begin{figure}
\begin{center}
\includegraphics[width=1\linewidth]{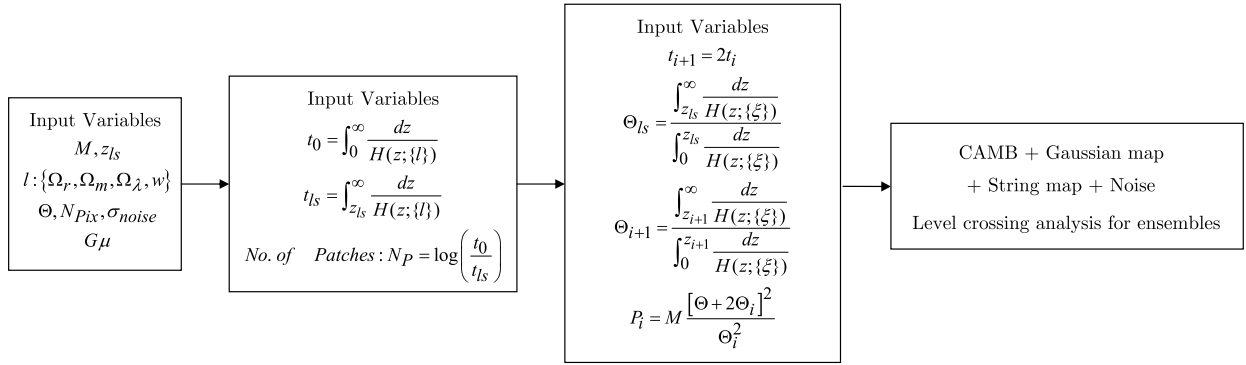}
\caption{\label{algorithm}Algorithm of map simulation and analysis used in this paper.}
\end{center}
\end{figure}
\section{Simulation of Mock CMB map}
In this section, we describe in detail our sate-of-the-art code
which has been written in Fortran language by authors to simulate
temperature fluctuations maps (also for more details see refs
\cite{brand080,brand081,brand09,pre93,pre93a,pre93b}). Generally,
our program has four parts: The first part creates pure Gaussian
fluctuations corresponding to the standard inflationary model in the
presence of various models governing the evolution of background. In
this paper we use $\Lambda$CDM model in
 the flat Universe. Nevertheless, our program can be easily modified to other cosmological models governing the background evolution of
 universe such as quintessence model and so on \cite{sadeghquin}. The second part  contains anisotropies produced by long cosmic strings by means
 of Kaiser-Stebbins effect. The superposition of Gaussian and cosmic strings anisotropies as well as the expected instrumental noise are
 produced in the third part. The normalization factor for each component will be computed in this part.
 Finally, the algorithm for searching the signatures of cosmic string relying on the level crossing analysis is given in the last part.
 To find more reliable results and reduced statistical errors, an ensemble averaging over at least 200 runs for a set of input variables
 will be done. As one can see in the following, we need a fine resolution map. However the random behavior of cosmic strings limits our
 simulation on the flat sky map approximation \cite{brand081,white99,seljak97}. In this case the Fourier modes are bases functions
 instead of being
 spherical harmonics. Before going further, for pedagogical purposes, we summarize all procedures to be done for map making in Figure (\ref{algorithm}).
\begin{figure}
\begin{center}
\includegraphics[width=0.5\linewidth]{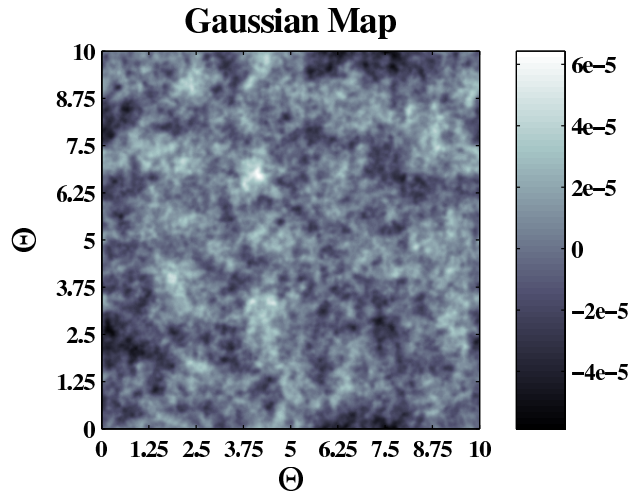}
\caption{\label{gaussian}A Gaussian map for $\Lambda$CDM model based on WMAP-7 observations. Map size is $10^{\circ}$ with $R=1.5'$.}
\end{center}
\end{figure}
\subsection{Gaussian map on the Flat sky}
In this subsection a Gaussian map for temperature fluctuations will
be generated. We use $\Lambda$CDM  model and the best fit values
have been inferred based on the most recent observations such as WMAP-7,
Supernova type Ia gold sample (SNIa) and Sloan Digital Sky Survey
(SDSS) with the most familiar initial power spectrum established by
standard inflationary scenario \cite{WMAP7,WMAP71}. As discussed in details
in Refs. \cite{brand080,brand081} and indicated in Figure
(\ref{algorithm}), first of all, we should determine the values of
initial parameters relevant to the Gaussian map. A part of these
parameter are related to the cosmological frame-work  and others are
the size of simulated map, $\Theta$, angular resolution, $R$, and
finally the variance of instrumental noise, namely $\sigma_{noise}$. Since we are interested in flat sky, namely a
simulated map with size less than $60^{\circ}$, two-dimensional
plane wave (Fourier basis functions) are used instead of spherical
harmonics.  The stochastic temperature fluctuations field on the flat sky will be written as:
\begin{eqnarray}\label{sim1}
\frac{\Delta T}{T}(k_x,k_y)&=&\sqrt{\frac{C_{l(k_x,k_y)}}{2}}\left[Z_1(k_x,k_y)+iZ_2(k_x,k_y)\right]
\end{eqnarray}
Actually to guaranty the Gaussianity of distribution based on central limit theorem, we add $Z_1(k_x,k_y)$ and $Z_2(k_x,k_y)$ as two independent Gaussian random numbers with zero mean and unit variance. The factor $2$ in the denominator stands for canceling the corresponding variance of Gaussian random numbers. To set the value of $C_l$ for a given set of $(k_x,k_y)$ in inverse degrees, one can compute $l$ as:
$l=\frac{360}{2\pi}\sqrt{k_x^2+k_y^2}$. The initial power spectrum,  $C_l$, is determined by running CAMB software\cite{brand081,CAMB}. The power spectrum for non-integer values  of $l$ can be evaluated by linear interpolation in the produced $C_l$. To this end we modify the public CAMB software for underlaying theoretical model and set the best fit values of corresponding cosmological parameters based on constraints given by  the most recent observations.
Finally to keep the statistical isotropy and to diminish the undesired preferred direction appearing in simulation, we construct four independent Gaussian maps and superimpose these separate components according to \cite{brand080}:
\begin{eqnarray}
&&\frac{\Delta T}{T}(x,y)_G=\nonumber\\
&&\frac{1}{2}\left[\frac{\Delta T}{T}(x,y)_{G1}+\frac{\Delta T}{T}(x_{\rm max}-x,y)_{G2}+\frac{\Delta T}{T}(x,y_{\rm max}-y)_{G3}+ \frac{\Delta T}{T}(x_{\rm max}-x,y_{\rm max}-y)_{G4}\right]
\end{eqnarray}
The pre-factor $1/2$ is necessary to keep the initial standard deviation. As mentioned before, in this paper we simulate Gaussian
temperature fluctuations based on $\Lambda$CDM  \cite{WMAP7,WMAP71} including best fit values for cosmological parameters based on WMAP-7 observations.
Figure (\ref{gaussian}) shows a  typical simulated Gaussian map for $\Lambda$CDM  model.
\begin{figure}
\begin{center}
\includegraphics[width=0.5\linewidth]{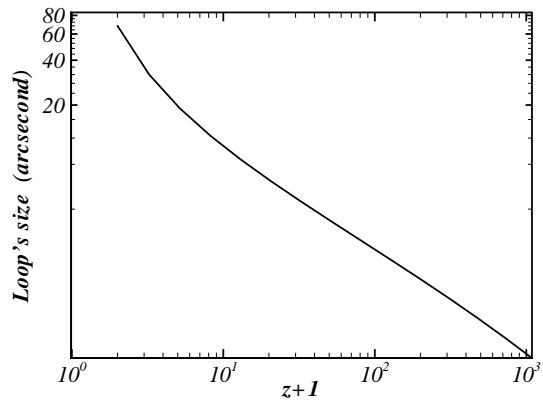}
\caption{\label{loop}Typical size of loop in arc-second units  as a function of redshift.}
\end{center}
\end{figure}
\subsection{Fluctuations Produced by Cosmic Strings}
In this subsection we explain in detail the algorithm used for
generating cosmic strings by using the Kaiser-Stebbins effect which
originally discussed by Perivolaropoulos \cite{pre93b} ( for more
details see Refs.
\cite{brand081,brand080,brand09,pre93,pre93b,pre93a}).  In our toy
model, since the size of a typical loop of string is about
$10^{-4}H^{-1}$ \cite{allen90,david88} (see Figure (\ref{loop})) and
the resolution of simulated map is of the order of one arc-minute so
we can ignore the contribution of these closed strings and suppose
that the main anisotropy in the string map is due to the straight
cosmic strings. In addition,  it has been shown that loops
contribute to temperature fluctuations via the well-known
Sachs-Wolfe effect and due to the small bandwidth of simulated map
the overall effects can be represented as Gaussian noise
\cite{sachs67,pre93a}.

\begin{figure}
\begin{center}
\includegraphics[width=0.7\linewidth]{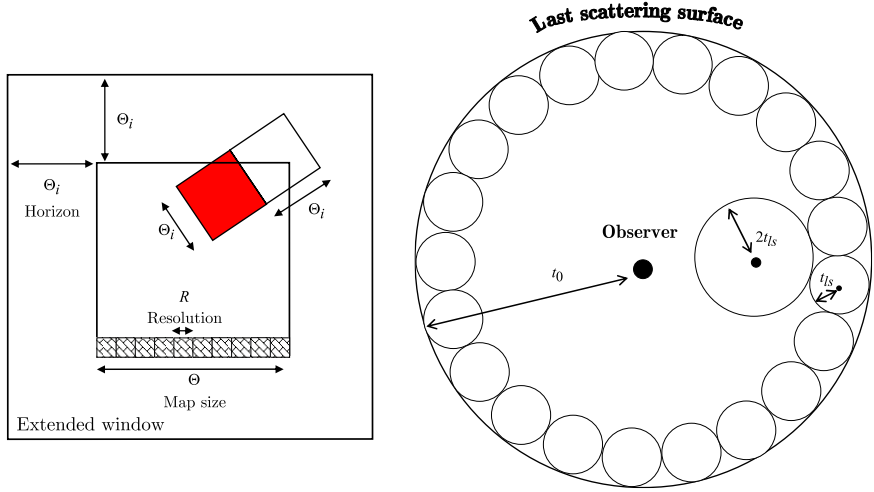} 
\caption{\label{successive1}Left panel shows a sketch of an extended window used to simulate the effect of one cosmic string on the CMB temperature of $\Theta\times\Theta$ map at redshift $z_{i}$. 
Generally, strings located in the extended map of size $(\Theta+2\Theta_i)^2$ can affect the simulated patch. Right panel corresponds to a schematic view of horizons at successive steps to accumulate KS effects on the
temperature fluctuations through traveling from the last scattering surface toward an observer. After about two times of the horizon length scale,
photons are encountered with a new Hubble volume, consequently, new string network makes kicks on them.}
\end{center}
\end{figure}

\begin{figure}
\begin{center}
\includegraphics[width=0.5\linewidth]{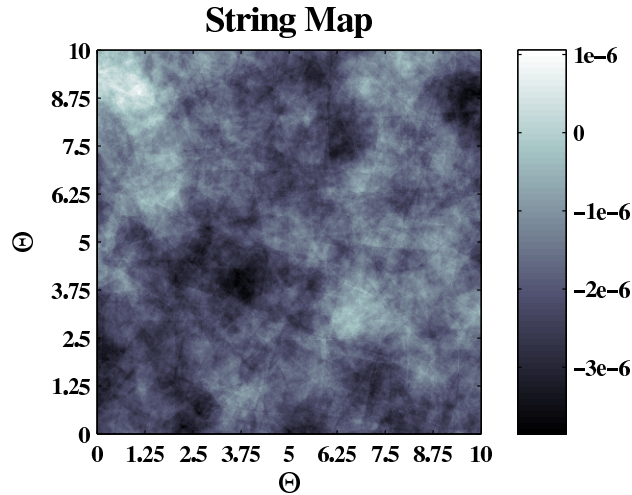}
\caption{\label{string1}A string map for $G\mu=1\times 10^{-7}$. Map
size equates to $10^{\circ}$ with $R=1.5'$.}
\end{center}
\end{figure}

The scaling behavior of the correlation length scale of
straight strings demonstrates that the number of strings
crossing a given Hubble volume to be fixed. In another word the
number density of strings to be invariant under rescaling of
correlation length scale
\cite{allen01,allen90,ander03,arms99,david88,land03,urr07}.
Therefore, one can deduce the numbers of  straight strings crossing
each Hubble volume at any given time to be  $M=10$. Their overall
properties such as orientations and velocities are statistically
uncorrelated for length scale larger than the corresponding horizon
scale \cite{david88,allen90,cope92}. The time interval in our
simulation is limited to the interval $[t_{ls},t_0]$  and since the
cosmic strings have relativistic velocity,  one can suppose that
after $2t_H$ ($t_H$ is Hubble time),  a new Hubble volume will be
generated and consequently a new network of cosmic strings affects
the propagation of the photons transmitted freely after last
scattering surface based on the Keiser-Stebbins phenomenon. This is
also justified regarding the fact that a photon ray travels most of
the Hubble volume during twice the Hubble time, and after that this ray 
encounters a new volume and will be affected by a new string network. So
one can suppose that at successive time intervals according to the
relation, $t_{i+1}=2t_i$,  the photon ray could be affected by a new
string network.  According to this statement, the number of separate
string networks (patches) to be simulated is given by (see Figure
(\ref{successive1})):
\begin{eqnarray}
N_P=\log_2\left(\frac{t_0}{t_{ls}}\right)\simeq 15
\end{eqnarray}
\begin{figure}
\begin{center}
\subfigure{\includegraphics[width=1\linewidth]{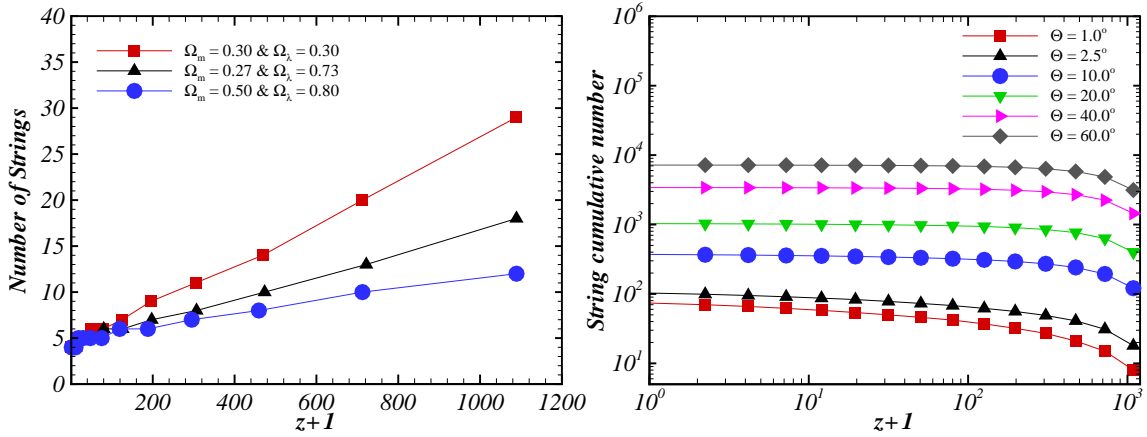}}
\caption{\label{cumulative}Left panel corresponds to the number of straight cosmic
strings in unit of $M$ ($M$ is the number of cosmic strings at each Hubble volume given by the scaling solution)
as a function of redshift for a simulated map with size of $2.5^{\circ}\times 2.5^{\circ}$. Right panel indicates
cumulative string number as a function of redshift for various values of simulated map size for a flat Universe with $\Omega_m=0.27$ and $\Omega_{\lambda}=0.73$.}
\end{center}
\end{figure}
In the string map for the recursion condition, $t_{i+1}=2t_i$, 
we let the new network of cosmic strings to kick the photon paths. The apparent angular size of
horizon at $z_i$ is given by:
\begin{equation}\label{sizehub}
\Theta_i=\frac{\int_{z_i}^{\infty}\frac{dz}{H(z;\{\xi\})} }{\frac{1}{H_0\sqrt{|{\Omega_K}|}}{\mathcal {F}}\left[\sqrt{|{\Omega_K}|}\int_{0}^{z_i}\frac{H_0dz}{H(z;\{\xi\})}\right]}
\end{equation}
here ${\mathcal{F}}(x)$ is $x$, $\sin(x)$ and $\sinh(x)$ for flat, closed and open universe, respectively.  In the above equation, $H(z;\{\xi\})$ is the Hubble  parameter and $\{\xi\}$ includes all model parameters.
The number of straight cosmic strings with size $\Theta_i$ at $z_i$ which should be simulated and placed randomly on the desired simulated map will be \cite{brand081,brand080,brand09,pre93}:
\begin{equation}
P_i=M\frac{\left[\Theta+2\Theta_i\right]^2}{\Theta_i^2}
\end{equation}
where $\Theta$ is the size of simulated map with resolution equal to
$R$, consequently the number of pixels for extended map at  $i$th
step of simulation is $[\Theta+2\Theta_i]/R$. Actually, to retain
all fluctuations produced by cosmic string in a limited area of sky
which has been simulated we establish an extended area for each
step. The size of mentioned extended window will be
$\Theta+2\Theta_i$ at $i$th patch and  our simulated map to be
considered at the center of this window. Left panel of figure (\ref{successive1})
shows this configuration for fluctuations produced by
Kaiser-Stebbins phenomena on the cosmic background radiation
fluctuations.  Upon the calculation of the size of horizon for $i$th successive step by using Eq. (\ref{sizehub}), we embed
a straight string and apply a shift for temperature fluctuations
according to Eq.(\ref{kseq1}). Since the mean value of fluctuations
in the simulated map has no physical significance, we add half of
$\left (\Delta T/T\right )_s$ to one side of string and subtract the remaining
value from other side, namely we use
\begin{equation}\label{kseq3}
\left(\frac{\Delta T}{T}\right)_s=\pm4\pi G\mu \gamma_s v_s r
\end{equation}
where $r\equiv |\hat{\textbf{n}}.(
\hat{\textbf{v}}_s\times\hat{\textbf{e}}_s)|$. The direction of
observer for a small field in the map is approximately constant.
Since the string's velocities and orientations are random  and
furthermore  $r$ takes into account the contribution of orientation
and projection effect of cosmic string, consequently its value will
be given by generating a uniform random number over the interval
$[0,1]$. In addition a binary flag is produced to decide which side
of string to be positive and which one will be negative for
temperature fluctuations. The location of string in the extended
window is also demonstrated by an additional proper random number.
We take into account the contribution of projection from
3-dimensions to 2-dimensions of cosmic string by  multiplying the
proper size of strings at each patch by the cosine of the angle
deriving from $[0,\frac{\pi}{2}]$ interval randomly.  We fix the
relevant coefficient for string simulation e.g. the number of cosmic
string for each Hubble volume,  $M=10$ and $v_s\gamma_s=0.15$
\cite{alb85}. It must be pointed out that the fixed values for
string quantities change significantly from one theoretical model to
the other, consequently in this toy model, we take most popular
values for mentioned relevant quantities \cite{vilinkin00}. As shown
in Figure (\ref{cumulative}) the number of strings for each step
depends on the size of simulated map as well as the underlying
cosmological theory. Finally, temperature fluctuations simulated for
various patches should be superimposed to reach the final situations
for a given pixel in the simulated map. Figure (\ref{string1})
displays a string map for which the cosmic tension is $G\mu=1\times
10^{-7}$. As an
additional check for consistency of our simulation we compute the
power spectrum of our pure string simulated map for different values
of $G\mu$ as well as pure Gaussian generated map in small angle approximation  \cite{white99,mark09,spergel08}. 
\begin{figure}
\begin{center}
\includegraphics[width=0.6\linewidth]{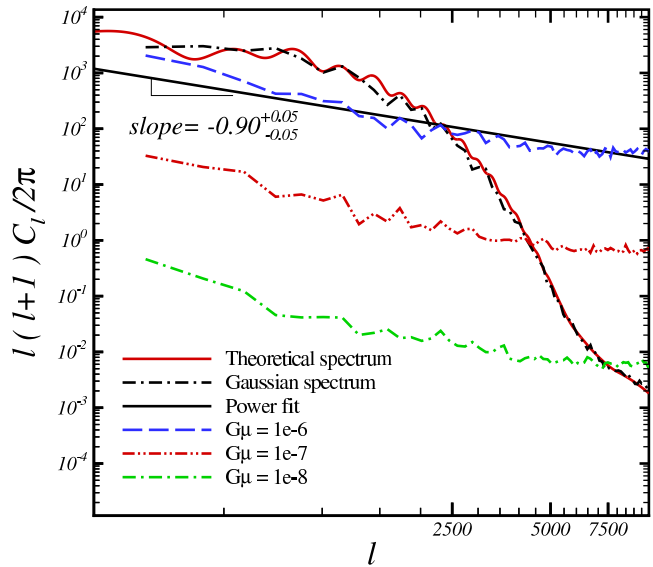}
\caption{\label{powerfig}Angular TT power spectrum of CMB
fluctuation. Solid line is calculated by CAMB software for the best
fit  values of $\Lambda$CDM based on 7-year WMAP data. Dashed line
corresponds to a simulated Gaussian map with size $2.5^{\circ}
\times 2.5^{\circ}$  and map resolution is $R=1'$. Long-dashed,
dashdd and dashdot lines correspond to fluctuations Long-dashed,
dashdd and dashdot lines correspond to fluctuations generated by
cosmic strings for some typical values of $G\mu$ indicated in the
plot. The power-law fitted function is also presented in figure with
the scaling exponent equates to $\eta=0.90_{-0.05}^{+0.05}$ for
$l\gg1$ \cite{spergel08}.}
\end{center}
\end{figure}
Figure (\ref{powerfig}) indicates the angular power spectrum for pure
Gaussian and pure cosmic string simulated maps for typical field of
view of $2.5^{\circ}$ and a resolution of $1.5'$. Since the size and
resolution of generated map is finite and temperature fluctuations
are limited to  the flat sky, the multiple moment will run from
$l\simeq10^2$ to $10^4$. In addition the expected scaling behavior
for pure cosmic string map satisfies the relation  $l(l+1)C_l\propto
l^{-\eta}$ for $l\gg 1$ with $\eta\simeq 0.90_{-0.05}^{+0.05}$.  Let
us explain the final task regarding the simulation of a Gaussian map
including induced cosmic string and instrumental noise. As discussed
before we simulate all components separately and according to the linear
perturbation theory, we superpose all sources to include all
phenomena and reach the final situation. 
We do not expect the observed temperature power spectrum to be changed in
the presence of cosmic string, consequently the multiplication by a
factor, $\omega$ less than unity causes the power spectrum to remain
unchanged. The instrumental noise does not modify the real sky map
and its significance is demonstrated by the maximum amplitude of
noise represented by $\frac{\Delta T}{T}\big{|}_{max}$. This maximum
will be taken as $\frac{\Delta T}{T}\big{|}_{max}=10\mu K$
throughout this paper according to the anticipated instrumental
noise in South Pole Telescope (SPT) \cite{SPT}. To finalize our
simulation, we use the same method explained
in Ref. \cite{brand081}. The power spectrum of simulated map can be read as follows:
\begin{equation}\label{summation4}
C_{l(G+S)}=\omega^2C_{l(G)}+C_{l(S)}
\end{equation}
The cross-correlated term is zero due to independency of Gaussian and string components.
We demand the left hand side of the above equation to be equal to
power spectrum derived from observation,
given by CAMB software. Therefore for each $l$ from $l_{min}$ to
$l_{max}$, we find a separate $\omega$'s. To compute a single
$\omega$, we rely on Bayesian statistics for observations,
$\{X\}:\{\omega(l)\}$ and model parameters, $\{\Omega\}$. We define \cite{fab04}:
 $\chi^2(\Omega)=\sum_{l=l_{min}}^{l=l_{max}}[\omega_{{\rm obs}}(l)-\omega_{{\rm the}}(l;\Omega)]^2$. 
Here $\omega_{{\rm obs}}(l)$ is computed directly from
Eq.(\ref{summation4}) and $\omega_{{\rm the}}(l;\Omega)$ derived by
an arbitrary merit function which is supposed to be a first order
polynomial function throughout this paper. 
\begin{figure}
\begin{center}
\includegraphics[width=1.0\linewidth]{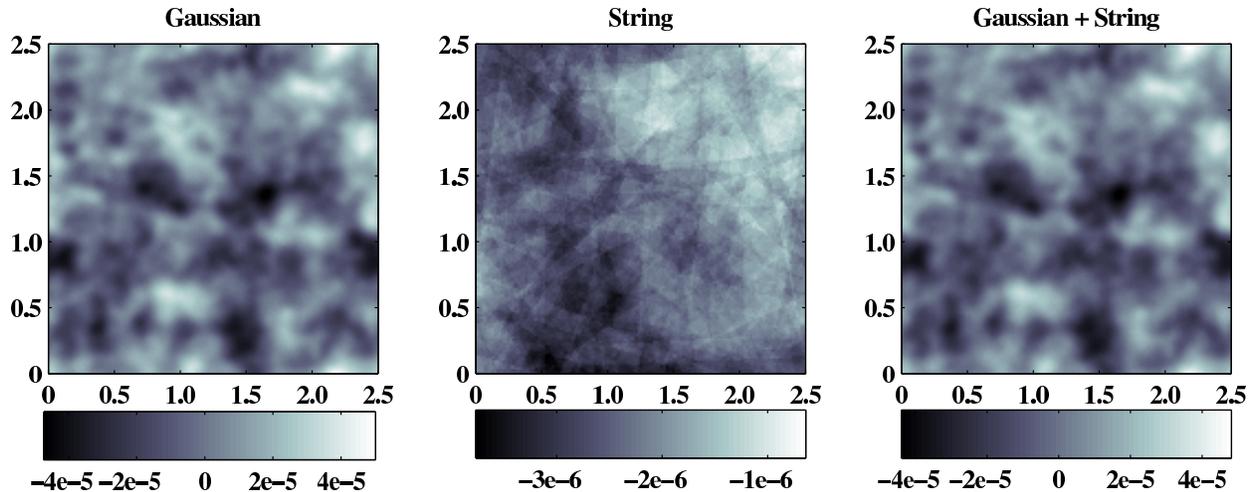}
\caption{\label{all1}Left panel shows the Gaussian map. The middle panel corresponds to pure cosmic
 string for $G\mu=1\times 10^{-7}$. The superposition of all components is shown in right panel. The resolution of these map is $R=1'$.}
\end{center}
\end{figure}
\begin{figure}
\begin{center}	
\includegraphics[width=1.0\linewidth]{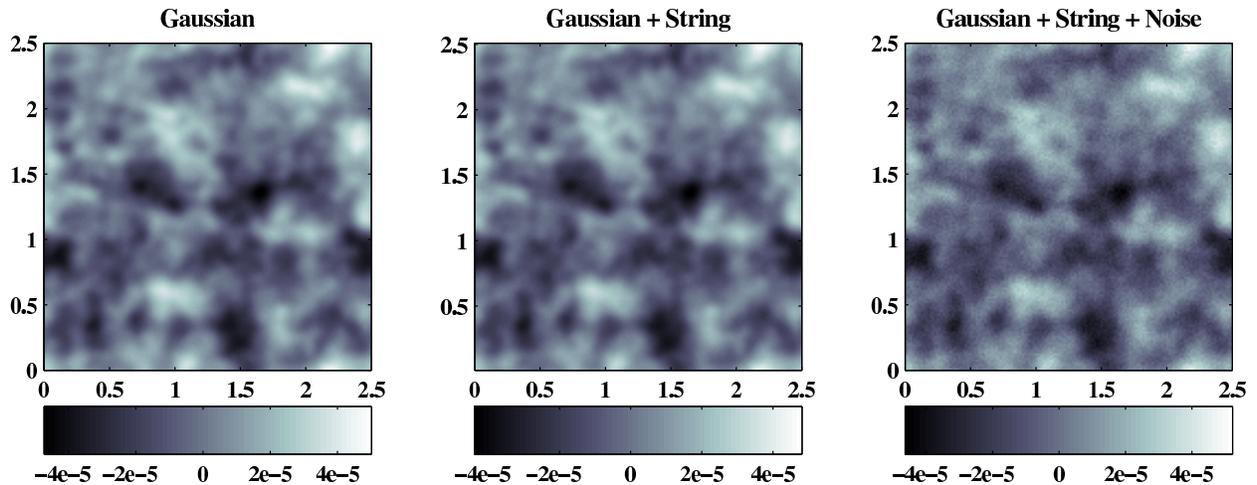}
\caption{\label{all2}Left panel shows the Gaussian map. The middle
one corresponds to Gaussian with cosmic strings added for
$G\mu=1\times 10^{-7}$. The superposition of  all components with
instrumental noise is shown in the right panel. Maximum value for
temperature fluctuations created by noise is considered to be
$\left(\frac{\Delta T}{T}_{max}\right)_N=10\mu K$. The size of all
maps is $2.5^{\circ}$ with a resolution of $R=1'$.}
\end{center}
\end{figure}
The y-intercept of merit function will be assumed as a best fit
value of $\omega$ \cite{brand081}. It must be pointed out that to
improve the reliability of $\omega$, we do various simulations and
finally by ensemble averaging, an averaged value for $\omega$ can be
retrieved. Figure (\ref{all1}) indicates different components
simulated without expected instrumental noise. Figure (\ref{all2})
shows simulated map including instrumental noise.  Now every things
ready to investigate temperature fluctuations on the flat sky with
and without cosmic strings.  We rely on a robust statistical method
namely, level crossing analysis to explore the capability of our
method to detect the cosmic strings for various values of string's
tensions.
\begin{figure}
\begin{center}
\includegraphics[width=1.0\linewidth]{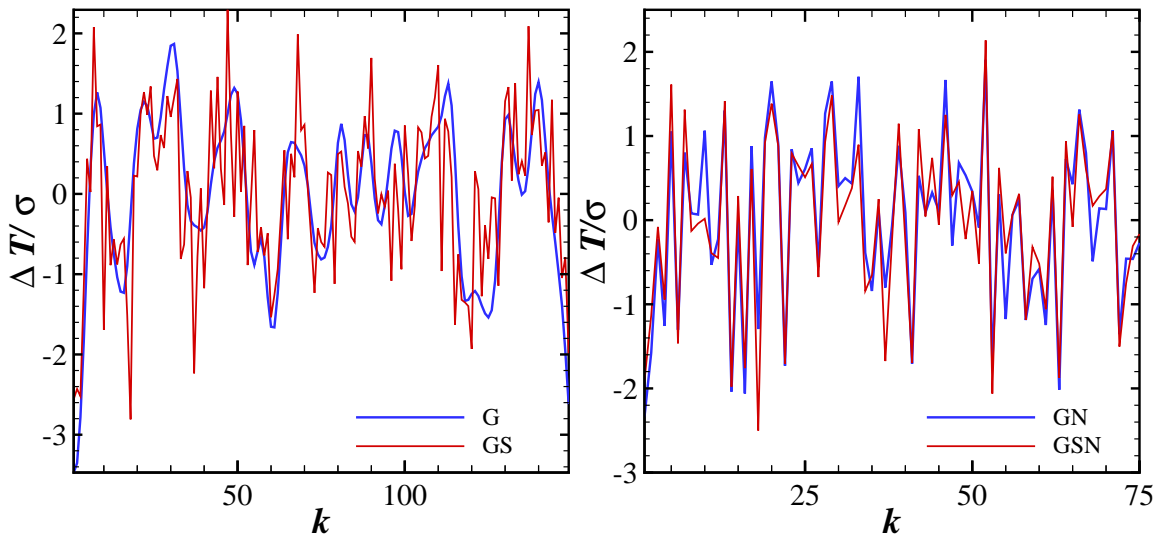}
\caption{\label{level_temp}Left panel shows the $1+1$-dimensional increment of temperature fluctuations for Gaussian and superimposed cosmic string component with $G\mu=2\times 10^{-7}$. Right panel corresponds to the same as left panel except including instrumental noise.}
\end{center}
\end{figure}
\section{Analysis of CMB Map}
\begin{figure}
\begin{center}
\includegraphics[width=1.0\linewidth]{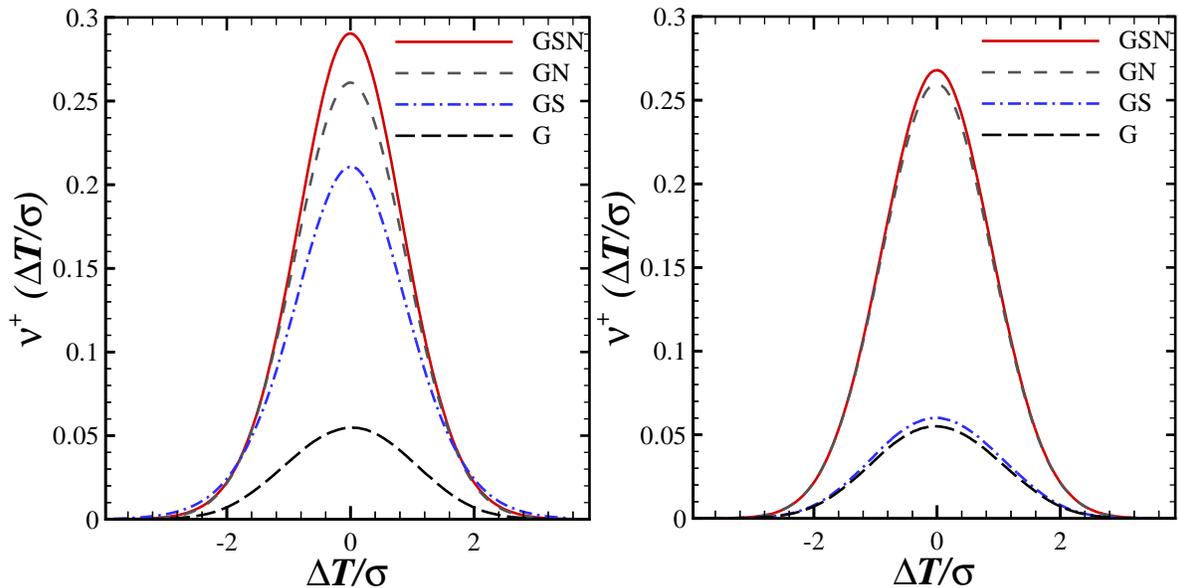}
\caption{\label{level20} Level crossing for increment of temperature fluctuations for various simulated components. The value of $G\mu$ for left and right panel are $1\times 10^{-6}$ and $1\times 10^{-7}$, respectively.}
\end{center}
\end{figure}

\begin{figure}
\begin{center}
\includegraphics[width=1.0\linewidth]{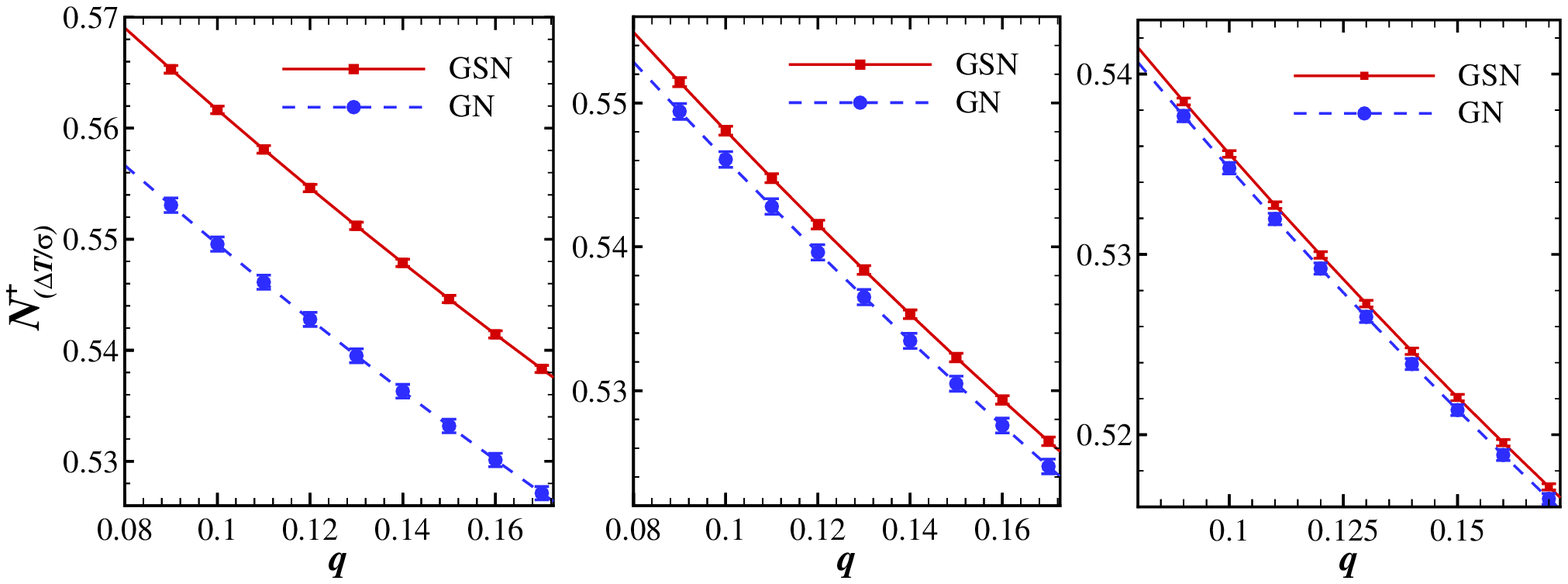}
\caption{\label{nq} Generalized roughness function versus moment,
$q$,  for Gaussian$+$ String $+$ Noise and Gaussian $+$ Noise. The
size of symbols are almost equal to their error bars. Here we run
over 200 ensembles. The value of string's tension from left to right are $G\mu=1\times 10^{-7}$, $G\mu=1\times 10^{-8}$ and $G\mu=6\times 10^{-9}$, respectively.}
\end{center}
\end{figure}

\begin{figure}
\begin{center}
\includegraphics[width=0.6\linewidth]{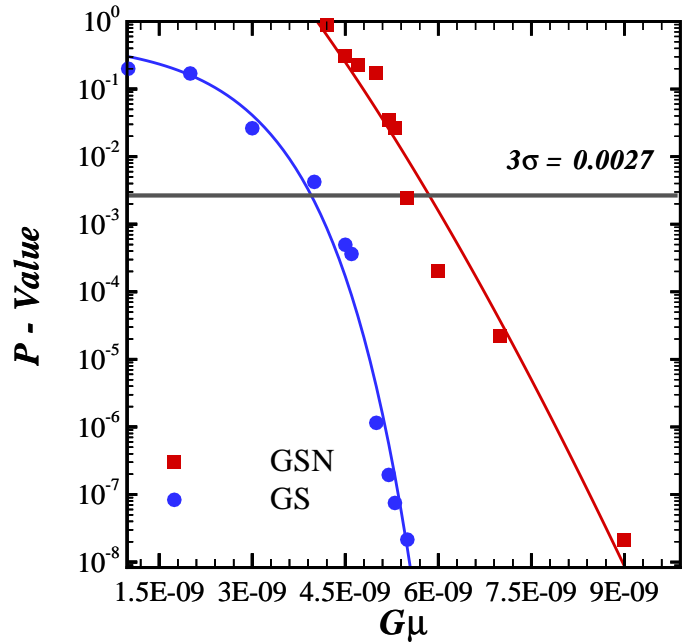}
\caption{\label{pvalue} P-value as a function of $G\mu$. We run over
200 ensembles to compute each point of this plot. To make more obvious we fitted data by typical fitting functions. }
\end{center}
\end{figure}

As explained in details in section II, we are interested in upcrossings the temperature fluctuations at an arbitrary level, $\alpha$. Or according to probability point of view, we should compute the volume of joint probability density function which is satisfied in the proper conditions mentioned in section II (see Figure (\ref{proba})). We expect the fluctuations in the presence of cosmic strings to be rougher than pure Gaussian fluctuations. Since the level crossing analysis is applicable for one-dimensional series we construct $1+1$-dimensional series from simulated map. Furthermore due to the statistical isotropy property of simulated map \cite{amir1,amir2,sadeghaniso}, the $1+1$-dimensional data sets will be created in all given directions as different ensembles for analysis. Actually these series contain temperature fluctuations as a function of pixels in one dimension in an arbitrary direction. Hereafter  we turn to investigate $1+1$-dimensional signals as an input for level crossing subroutine which is the final part of  the main simulation program. We introduce the increment of fluctuation as:
\begin{equation}
\Delta T(k)\equiv\left( \frac{T(k)-\bar{T}}{\bar{T}}\right)-\left (\frac{T(k-1)-\bar{T}}{\bar{T}}\right)
\end{equation}
In Figure (\ref{level_temp}) we have plotted a typical
$1+1$-dimensional increment series for Gaussian and cosmic string
signals embedded in Gaussian fluctuations. We see that in the
presence of strings, fluctuations near the mean as well as far from
the mean have been changed. The magnitude of these changes depends
on value of cosmic string tension. As pointed out before, we
enumerate the number of crossings with positive slope at whole
existence level for all $1+1$-dimensional data set in a map for a
given $G\mu$. It is worth noting that according to conservation law,
for long-run simulation all crossings with negative slope are
statistically equivalent to that of with positive slope
\cite{percy00}. Based on this analysis, we find many quantities which
may be potentially used as a criterion to distinguish between the
maps with and without straight cosmic strings. In Figure
(\ref{level20}) we plot the $\nu_{\alpha}^+$ versus $\alpha$ for
various values of $G\mu$. Generally, total crossings in the Gaussian
map is less than that of with cosmic string.  As we expect, the
existence of instrumental noise decreases the efficiency of our
analysis.  To compare simulated maps we refer to  the generalized
form of roughness of signals which is given by Eq.(\ref{ntq}),
$N_{tot}^+(q)$. The various values  for $q$'s take into account
crossings with different measures in the total level crossings.
Figure (\ref{nq}) indicates the generalized roughness function for
some typical values of $G\mu$. To compare the value of generalized
roughness function we apply Student's t-test for equal sample size
and unequal mean and standard deviation for each $q$'s which is
defined by:
\begin{equation}
t(q)=\left(\overline{N_{tot}^{+}}(\oplus,q)-\overline{N_{tot}^{+}}(\otimes,q)\right)\times \sqrt{\frac{N_{run}}{\sigma_{\oplus}^2(q)+\sigma_{\otimes}^2(q)}}
\end{equation}
here the symbols $\oplus$ and $\otimes$ stand for $Gaussian$ (G) and
$Gaussian+String$ (GS) for simulated map without noise,
respectively. For simulation including instrumental noise these
symbols  are replaced by $Gaussian+Noise$ (GN) and
$Gaussian+String+Noise$ (GSN), respectively. The
$\overline{N_{tot}^{+}}$ is the ensemble average of generalized
roughness function for each $q$'s. The corresponding standard
deviation is given by $\sigma(q)$. $N_{run}$ indicates the number of
simulated ensembles. The P-value, $p(q)$ for $t(q)$ based on
t-distribution function with $2N_{run}-2$ degrees of freedom, will
be determined. By the combination of above P-values we introduce the
following quantity:
\begin{equation}
\chi^2=-2\sum_{q=q_{min}}^{q_{max}}\ln p(q)
\end{equation}
Finally, the last P-value, $P_{final}$ associating to the $\chi^2$
by using chi-square distribution function for
$2\left(\frac{q_{max}-q_{min}}{\Delta q}\right)-2$ degrees of
freedom will be computed. For $3\sigma$ significance level namely
$P_{final}<0.0027$, we can conservatively say that there exists a
significance difference. According to the value of $P_{final}$
indicated in Figure (\ref{pvalue}), we conclude that our method can
detect the cosmic string with $G\mu\gtrsim 4\times 10^{-9}$ and in
the presence of anticipated instrumental noise the lower bound is
$G\mu\gtrsim 5.8\times 10^{-9}$. The final strategy for detecting
cosmic string is as follows: At first by using a map from
observation we compute $\nu_{\alpha}^+$ and $N_{tot}^+(q)$. Second
we simulate a pure Gaussian map with expected instrumental noise and
again the level crossing and generalized roughness functions to be
calculated. The P-value for deciding about the existence of
significant difference between $N_{tot}^+(q)$ for real and simulated
Gaussian maps will be determined. In the case of significant
difference, we change the value of $G\mu$ for simulated Gaussian map
including cosmic strings and compute the corresponding $P_{final}$.
Upon the $P_{final}$ for simulation to be equal to that of given for
observation we conservatively can express the existence of cosmic
strings in the observed map with mentioned $G\mu$.
\section{Summary and Conclusions}
In this paper we relied on a robust method in complex system which
has been introduced originally to investigate the statistical
properties of $2+1$-dimensional surface by mapping to
$1+1$-dimensional data sets
\cite{percy00,movahed06,movahed05,movahed07,tabar02,tabar03}.
Using this method we explore the capability of finding the
footprints of straight cosmic string based on the KS effect on the
cosmic microwave background fluctuations since the last scattering
epoch up to now. To this end we take a simple toy model to simulate
the effect of cosmic strings demonstrated in Refs.
\cite{pre93,pre93b}. The contribution of loops in this model can be
ignored (see Figure (\ref{loop})). The superposition of kicks due to
the embedded cosmic strings cause some additional statistical
fluctuations in simulated map. Consequently  there exists a
difference between the number of level crossing for whole reachable
level in $1+1$-dimensional fluctuations of pure Gaussian map in
comparison with map containing cosmic strings. To quantify this
deviation we introduced an increment signal and use robust quantity
to take into account all deviations relative to pure Gaussian
fluctuations, namely generalized roughness function. For $q=0$ the
generalized roughness function corresponds to sum of  upcrossings at
all fluctuation levels which is a measure of roughness of underlying
signals. Based on joint probability density function of signal,
$p(y=\alpha,y')$, we found that the fluctuations near the mean
fluctuations could be a proper criterion to discriminate pure
Gaussian map from the map induced by cosmic strings. Our result
shows that level crossing analysis could place a bound for cosmic
string tension $G\mu\gtrsim 4\times 10^{-9}$ for 200 simulated maps
with $R=1'$.  To explore the contribution of instrumental noise we
relied on South Pole Telescope experiment and took
$\left(\frac{\Delta T}{T}\big{|}_{max}\right)_N=10\mu K$ for noise
map. The new bound for map including instrumental noise is
$G\mu\gtrsim 5.8 \times 10^{-9} $. The small change in the bound on
$G\mu$ due to the noise could be justified regarding Figure
(\ref{level20}). The sensitive part of $\nu_{\alpha}$ is crossing
near the mean value so the presence the undesired noise can only
have minor statistical effects on the ability of this model to place
a lower bound on the string's tension for small value of $G\mu$. In addition, our results for
both simulations were not affected by increasing the size of
ensemble up to 200 corresponding to increase the coverage of sky
map. For the realistic observations there are additional problems to
be addressed. The first one is the foreground and the instrumental
noise which is supposed to be considered by adding expected noise
and smoothing our simulation. As mentioned in Refs. \cite{brand081,brand080,brand09}
discontinuity related to the scanning strategy also puts extra
imprint on data set. It must point out that our method is sensitive
to overall fluctuations in all directions, therefore, since it is
expected that by increasing the coverage of underlying map the
statistics of fluctuations near the mean to be significant, the
mentioned spurious phenomenon will have a negligible contribution to
the final result. However we can remove all parallel discontinuities
in preferred direction imposed by scanning maneuvering of
observational instrument \cite{brand081}.
Final remark is that we would like to use more realistic models
\cite{land03,spergel08,shellard10} to simulate a map  taking all
contributions of cosmic strings on the cosmic microwave background
into account and implement our method to explore the effect of
cosmic strings in our future works.  In addition as a future study,
it is useful to apply this method to distinguish the effect of
generic cosmic strings and superstrings on the temperature
fluctuations.
The n-point peak-peak correlation function is also another method which could be used for the same purpose.

\section*{Acknowledgements}
We would like to thank H. Firouzjahi and H. Mos'hafi for useful comments and reading manuscript. Also we thank the anonymous referee for his/her useful recommendations  and comment regarding the peak-peak correlation function method. M. Sadegh Movahed is grateful to the school of astronomy (IPM) for their hospitality.
This work has been partly supported by IPM.

\end{document}